# RF SURFACE IMPEDANCE OF A TWO-BAND SUPERCONDUCTOR, CONSIDERING MAGNESIUM DIBORIDE FOR ACCELERATOR APPLICATIONS

Binping Xiao[†]
*Collider-Accelerator Department, Brookhaven National Laboratory, Upton, New York 11973-5000, USA*
C. E. Reece
*Thomas Jefferson National Accelerator Facility, Newport News, Virginia 23606, USA*

In this paper, we apply the two-band extension [1] of BCS theory [2] into the Mattis-Bardeen theory [3] to obtain the surface impedance of a two-band superconductor, and apply it to magnesium diboride ($MgB_2$) for radiofrequency (RF) superconductivity applications. The numerical results for $MgB_2$ are in good agreement with the previously published experimental results [4]. The surface impedance properties are clearly dominated by the smaller gap, significantly limiting utility in the 10-20 K regime that might otherwise have been attractive.

## I. Introduction

The recent activities of superconducting radiofrequency (SRF) accelerating cavities for the Linac Coherent Light Source II (LCLS-II) particle accelerator made from bulk niobium (Nb) materials are the state-of-art activities for high efficiency CW accelerators for exploring frontier physics. The highest quality factor of single cell cavities was pushed to $7\times10^{10}$ at 1.3 GHz and 2 K temperature [5, 6], essentially fully exploiting the theoretical potential of niobium to accelerating field gradients of at least 20 MV/m [7]. Alternative superconducting materials with critical temperature and superheating critical field higher than those of Nb are of great interest as the next opportunity for continued technological progress. $MgB_2$ has been one of the candidate materials under investigation for possible future use in SRF applications since its discovery to be a superconductor by Nagamatsu *et al*. in 2001 [8]. $MgB_2$ is considered to be a conventional superconductor with a high critical temperature, $T_c$, of ~39-40 K [9]. It has two energy gaps, with a $\pi$-gap at 2.3 meV and a $\sigma$-gap at 7.1 meV [10], two corresponding coherence lengths between 1.6 and 5.0 nm and between 3.7 and 12.8 nm[9], and penetration depth between 85 and 203 nm [9]. At a temperature much lower than its critical temperature, the surface resistance is considered to be dominated by the $\pi$-gap, and could potentially be much lower than that of Nb considering that Nb's energy gap is 1.5 meV, and its critical temperature is 9.27 K. This feature also makes $MgB_2$ attractive for use in high-performance multilayer film coatings proposed by Gurevich [11]. Varies tests have been made to measure the surface resistance and reactance of $MgB_2$ [4, 12-14]. In reference [4], the authors used the Surface Impedance Characterization (SIC) system [15] at Jefferson Lab (JLab) to test SRF properties of $MgB_2$ at 7.4 GHz, with temperature ranges from 2.2 K to critical temperature. This work was distinguished from the previous works by higher resolution [4].

Mattis and Bardeen (MB) [3], and Abrikosov, Gor'kov and Khalatnikov (AGK) [16], independently derived the theory to calculate the surface impedance of single band BCS superconductors. These two theories have the same expression in the low field limit. Computer codes developed by J. Halbritter based on AGK and the one developed by J. P. Turneaure based on MB are routinely used to calculate the Nb surface impedance for SRF applications [17]. However, there is yet a lack of a treatment of the surface impedance of two-band superconductors based on MB or AGK. The two-band extension by Suhl *et al*. [1] was adopted to explain the energy gaps of $MgB_2$ at different temperatures[10, 18-20]. In this paper, we further apply it into MB theory to obtain the surface impedance of $MgB_2$ and compare the simulated results with measured data [4].

## II. RF surface impedance of a two-band superconductor

---

[†]binping@bnl.gov

Following BCS theory [2], with temperature $T$ close to 0, the probability of the state ($\mathbf{k}\uparrow$, $-\mathbf{k}\downarrow$) being occupied by a pair of particles is $h_k$. With a finite $T$, single electrons start to appear. $f_k$ is defined as the probability that state $\mathbf{k}\uparrow$ or $-\mathbf{k}\downarrow$, with Bloch energy relative to the Fermi sea of $\varepsilon_k$, being occupied. Similar to expression (1) in reference [1], for a two-band superconductor, the total free energy of a superconductor in the superconducting state can be expressed as:

$$F = 2\sum_k |\varepsilon_{ks}|[f_{ks} + (1-2f_{ks})h_{ks}(|\varepsilon_{ks}|)] + 2\sum_k |\varepsilon_{kd}|[f_{kd} + (1-2f_{kd})h_{kd}(|\varepsilon_{kd}|)] - V_{ss}\sum_{kk'}[h_{ks}(1-h_{ks})h_{k's}(1-h_{k's})]^{\frac{1}{2}}(1-2f_{ks})(1-2f_{k's}) - V_{dd}\sum_{kk'}[h_{kd}(1-h_{kd})h_{k'd}(1-h_{k'd})]^{\frac{1}{2}}(1-2f_{kd})(1-2f_{k'd}) - V_{sd}\sum_{kk'}[h_{ks}(1-h_{ks})h_{k'd}(1-h_{k'd})]^{\frac{1}{2}}(1-2f_{ks})(1-2f_{k'd}) - V_{ds}\sum_{kk'}[h_{kd}(1-h_{kd})h_{k's}(1-h_{k's})]^{1/2}(1-2f_{kd})(1-2f_{k's}) + 2kT\sum_k[f_{ks}\ln f_{ks} + (1-f_{ks})\ln(1-f_{ks}) + f_{kd}\ln f_{kd} + (1-f_{kd})\ln(1-f_{kd})] \quad (1)$$

The subscripts $d$ and $s$ represents different bands in the superconductor. $V_{ss}$, $V_{dd}$, $V_{sd}$ and $V_{ds}$ are the averaged interaction energies resulting from phonon emission and absorption by $s$-$s$, $d$-$d$, $s$-$d$ and $d$-$s$ processes, minus the corresponding shielded Coulomb interaction terms [1]. By minimizing $F$ with respect to $h_{ks}$, $h_{kd}$ and $f_{ks}$, $f_{kd}$ separately, we obtain:

$$h_{ks} = \frac{1}{2}\left(1 - \frac{\varepsilon_{ks}}{E_{ks}}\right), \quad h_{kd} = \frac{1}{2}\left(1 - \frac{\varepsilon_{kd}}{E_{kd}}\right), \quad (2)$$

$$f_{ks} = f(E_{ks}), \quad f_{kd} = f(E_{kd}), \quad (3)$$

with

$$f(E) = \frac{1}{e^{\beta E}+1}, \quad \beta = \frac{1}{kT} \quad (4)$$

$$D = \sum_k [h_{kd}(1-h_{kd})]^{1/2}(1-2f_{kd}),$$
$$S = \sum_k [h_{ks}(1-h_{ks})]^{1/2}(1-2f_{ks}) \quad (5)$$

$$A = V_{ss}S + V_{sd}D = 2\varepsilon_{ks}\sqrt{h_{ks}(1-h_{ks})}/(1-2h_{ks}),$$
$$B = V_{ds}S + V_{dd}D = 2\varepsilon_{kd}\sqrt{h_{kd}(1-h_{kd})}/(1-2h_{kd}) \quad (6)$$

$$E_{ks} = [\varepsilon_{ks}^2 + A^2]^{\frac{1}{2}}, \quad E_{kd} = [\varepsilon_{kd}^2 + B^2]^{\frac{1}{2}} \quad (7)$$

The above expressions agree with the results in reference [1].

In order to calculate the SRF BCS surface impedance, one may start with the matrix elements of a single-particle scattering operator as in references [2, 3], using the two-band distribution equations (1-7) above.

The scattering between $ks$ and $k's$, as well as the scattering between $kd$ and $k'd$, should be the same as the case with single band gap, with energy gap $A$ for band $s$, and $B$ for band $d$.

To calculate the effect brought by the scattering between $kd$ and $k's$, and also between $ks$ and $k'd$, we list the matrix elements in Table I. In these columns, $h$, $h'$, $f$, $f'$, $E$, $E'$ are used to simplify the expressions of $h_k$, $h_{k'}$, $f_k$, $f_{k'}$, $E_k$ and $E_{k'}$, respectively. One may refer to references [2, 3] for more detail about this table.

TABLE I. Matrix elements of single particle scattering operator (between $ks$ and $k'd$).

| Wave functions | | | | Ground(+) or excited(-) | | Energy difference $W_i - W_f$ | Probability of initial state | Matrix elements | |
|---|---|---|---|---|---|---|---|---|---|
| Initial, $\psi_i$ | | Final, $\psi_f$ | | | | | | $c_{k's\uparrow}{}^* c_{kd\uparrow}$ or $c_{-k's\downarrow}{}^* c_{kd\uparrow}$ | $c_{-kd\downarrow}{}^* c_{-k's\downarrow}$ or $-c_{-kd\downarrow}{}^* c_{k's\uparrow}$ |
| $ks$ | $k'd$ | $ks$ | $k'd$ | $ks$ | $k'd$ | | | | |
| (a) | | | | + | + | $E_s - E_{d'}$ | $1/2 s_s(1-s_{d'}-p_{d'})$ | $[(1-h_s)(1-h_{d'})]^{1/2}$ | $-(h_s h_{d'})^{1/2}$ |
| X0 00 | | 00 X0 | | - | - | $-E_s + E_{d'}$ | $1/2 s_s p_{d'}$ | $(h_s h_{d'})^{1/2}$ | $-[(1-h_s)(1-h_{d'})]^{1/2}$ |
| X0 XX | | XX X0 | | + | - | $E_s + E_{d'}$ | $1/2 s_s p_{d'}$ | $-[(1-h_s)h_{d'}]^{1/2}$ | $-[h_s(1-h_{d'})]^{1/2}$ |
| | | | | - | + | $-E_s - E_{d'}$ | $1/2 s_s(1-s_{d'}-p_{d'})$ | $-[h_s(1-h_{d'})]^{1/2}$ | $-[(1-h_s)h_{d'}]^{1/2}$ |
| (b) | | | | + | + | $-E_s + E_{d'}$ | $1/2 s_{d'}(1-s_s-p_s)$ | $(h_s h_{d'})^{1/2}$ | $-[(1-h_s)(1-h_{d'})]^{1/2}$ |
| XX 0X | | 0X XX | | - | - | $E_s - E_{d'}$ | $1/2 s_{d'} p_s$ | $[(1-h_s)(1-h_{d'})]^{1/2}$ | $-(h_s h_{d'})^{1/2}$ |
| 00 0X | | 0X 00 | | + | - | $-E_s - E_{d'}$ | $1/2 s_{d'}(1-s_s-p_s)$ | $[h_s(1-h_{d'})]^{1/2}$ | $[(1-h_s)h_{d'}]^{1/2}$ |
| | | | | - | + | $E_s + E_{d'}$ | $1/2 s_{d'} p_s$ | $[(1-h_s)h_{d'}]^{1/2}$ | $[h_s(1-h_{d'})]^{1/2}$ |
| (c) | | | | + | + | $E_s + E_{d'}$ | $1/4 s_s s_{d'}$ | $[(1-h_s)h_{d'}]^{1/2}$ | $[h_s(1-h_{d'})]^{1/2}$ |
| X0 0X | | 00 XX | | - | - | $-E_s - E_{d'}$ | $1/4 s_s s_{d'}$ | $-[h_s(1-h_{d'})]^{1/2}$ | $-[(1-h_s)h_{d'}]^{1/2}$ |
| | | XX 00 | | + | - | $E_s - E_{d'}$ | $1/4 s_s s_{d'}$ | $[(1-h_s)(1-h_{d'})]^{1/2}$ | $-(h_s h_{d'})^{1/2}$ |
| | | | | - | + | $-E_s - E_{d'}$ | $1/4 s_s s_{d'}$ | $-(h_s h_{d'})^{1/2}$ | $[(1-h_s)(1-h_{d'})]^{1/2}$ |
| (d) | | | | + | + | $-E_s - E_{d'}$ | $(1-s_s-p_s)(1-s_{d'}-p_{d'})$ | $[h_s(1-h_{d'})]^{1/2}$ | $[(1-h_s)h_{d'}]^{1/2}$ |
| XX 00 | | 0X X0 | | - | - | $E_s + E_{d'}$ | $p_s p_{d'}$ | $-[(1-h_s)h_{d'}]^{1/2}$ | $-[h_s(1-h_{d'})]^{1/2}$ |
| 00 XX | | | | + | - | $-E_s + E_{d'}$ | $(1-s_s-p_s)p_{d'}$ | $-(h_s h_{d'})^{1/2}$ | $[(1-h_s)(1-h_{d'})]^{1/2}$ |
| | | | | - | + | $E_s - E_{d'}$ | $p_s(1-s_{d'}-p_{d'})$ | $[(1-h_s)(1-h_{d'})]^{1/2}$ | $-(h_s h_{d'})^{1/2}$ |

While under RF field with angular frequency $\omega$, the photon energy $\hbar(\omega-is)$ should be inserted into either the initial or the final state in Table I. Here a small positive parameter $s$, which will be set equal to zero in the final expression, has been introduced to obtain the real and imaginary part of surface impedance [3].

Based on the above analysis, the single-particle scattering operator, shown as equation (3.2) in reference [3] may be rewritten as four separate items,

$$L_{ss}(\omega, \varepsilon_s, \varepsilon_s') = \frac{1}{4}\left(1 + \frac{\varepsilon_s \varepsilon_s' + A^2}{E_s E_s'}\right)\left(\frac{1}{E_s - E_s' + \hbar(\omega-is)} + \frac{1}{E_s - E_s' - \hbar(\omega-is)}\right)(f_s' - f_s) + \frac{1}{4}\left(1 - \frac{\varepsilon_s \varepsilon_s' + A^2}{E_s E_s'}\right)\left(\frac{1}{E_s + E_s' + \hbar(\omega-is)} + \frac{1}{E_s + E_s' - \hbar(\omega-is)}\right)(1 - f_s' - f_s)$$

$$L_{dd}(\omega, \varepsilon_d, \varepsilon_d') = \frac{1}{4}\left(1 + \frac{\varepsilon_d \varepsilon_d' + B^2}{E_d E_d'}\right)\left(\frac{1}{E_d - E_d' + \hbar(\omega-is)} + \frac{1}{E_d - E_d' - \hbar(\omega-is)}\right)(f_d' - f_d) + \frac{1}{4}\left(1 - \frac{\varepsilon_d \varepsilon_d' + B^2}{E_d E_d'}\right)\left(\frac{1}{E_d + E_d' + \hbar(\omega-is)} + \frac{1}{E_d + E_d' - \hbar(\omega-is)}\right)(1 - f_d' - f_d)$$

$$L_{sd}(\omega, \varepsilon_s, \varepsilon_d') = \frac{1}{4}\left(1 + \frac{\varepsilon_s \varepsilon_d' + AB}{E_s E_d'}\right)\left(\frac{1}{E_s - E_d' + \hbar(\omega-is)} + \frac{1}{E_s - E_d' - \hbar(\omega-is)}\right)(f_d' - f_s) + \frac{1}{4}\left(1 - \frac{\varepsilon_s \varepsilon_d' + AB}{E_s E_d'}\right)\left(\frac{1}{E_s + E_d' + \hbar(\omega-is)} + \frac{1}{E_s + E_d' - \hbar(\omega-is)}\right)(1 - f_d' - f_s)$$

$$L_{ds}(\omega, \varepsilon_s', \varepsilon_d) = \frac{1}{4}\left(1 + \frac{\varepsilon_d \varepsilon_s' + AB}{E_d E_s'}\right)\left(\frac{1}{E_d - E_s' + \hbar(\omega-is)} + \frac{1}{E_d - E_s' - \hbar(\omega-is)}\right)(f_s' - f_d) + \frac{1}{4}\left(1 - \frac{\varepsilon_d \varepsilon_s' + AB}{E_d E_s'}\right)\left(\frac{1}{E_d + E_s' + \hbar(\omega-is)} + \frac{1}{E_d + E_s' - \hbar(\omega-is)}\right)(1 - f_s' - f_d) \quad (8)$$

For the integration purpose, expression (8) can be rearranged as,

$$L_{ss}(\omega, \varepsilon_s, \varepsilon_s') = -\frac{1}{2}(1 - 2f_s)\left\{\frac{E_s + \hbar(\omega-is) + \frac{\varepsilon_s \varepsilon_s' + A^2}{E_s}}{E_s'^2 - [E_s + \hbar(\omega-is)]^2} + \frac{E_s - \hbar(\omega-is) + \frac{\varepsilon_s \varepsilon_s' + A^2}{E_s}}{E_s'^2 - [E_s - \hbar(\omega-is)]^2}\right\}$$

$$L_{dd}(\omega, \varepsilon_d, \varepsilon_d') = -\frac{1}{2}(1 - 2f_d)\left\{\frac{E_d + \hbar(\omega-is) + \frac{\varepsilon_d \varepsilon_d' + B^2}{E_d}}{E_d'^2 - [E_d + \hbar(\omega-is)]^2} + \frac{E_d - \hbar(\omega-is) + \frac{\varepsilon_d \varepsilon_d' + B^2}{E_d}}{E_d'^2 - [E_d - \hbar(\omega-is)]^2}\right\}$$

$$L_{sd}(\omega, \varepsilon_s, \varepsilon_d') = -\frac{1}{2}(1 - 2f_s)\left\{\frac{E_s + \hbar(\omega-is) + \frac{\varepsilon_s \varepsilon_d' + AB}{E_s}}{E_d'^2 - [E_s + \hbar(\omega-is)]^2} + \frac{E_s - \hbar(\omega-is) + \frac{\varepsilon_s \varepsilon_d' + AB}{E_s}}{E_d'^2 - [E_s - \hbar(\omega-is)]^2}\right\}$$

$$L_{ds}(\omega, \varepsilon_d, \varepsilon_s') = -\frac{1}{2}(1 - 2f_d)\left\{\frac{E_d + \hbar(\omega-is) + \frac{\varepsilon_d \varepsilon_s' + AB}{E_d}}{E_s'^2 - [E_d + \hbar(\omega-is)]^2} + \frac{E_d - \hbar(\omega-is) + \frac{\varepsilon_d \varepsilon_s' + AB}{E_d}}{E_s'^2 - [E_d - \hbar(\omega-is)]^2}\right\}$$

and

$$I(\omega, R, T) = V_{ss} N_s N_s \int_{-\infty}^{\infty}\int_{-\infty}^{\infty}\left[L_{ss}(\omega, \varepsilon_s, \varepsilon_s') - \frac{f(\varepsilon_s) - f(\varepsilon_s')}{\varepsilon_s' - \varepsilon_s}\right] \times \cos[\alpha(\varepsilon_s - \varepsilon_s')] d\varepsilon_s' d\varepsilon_s + V_{dd} N_d N_d \int_{-\infty}^{\infty}\int_{-\infty}^{\infty}\left[L_{dd}(\omega, \varepsilon_d, \varepsilon_d') - \frac{f(\varepsilon_d) - f(\varepsilon_d')}{\varepsilon_d' - \varepsilon_d}\right] \times \cos[\alpha(\varepsilon_d - \varepsilon_d')] d\varepsilon_d' d\varepsilon_d + \frac{V_{sd} + V_{ds}}{2} N_s N_d \int_{-\infty}^{\infty}\int_{-\infty}^{\infty}\left[L_{sd}(\omega, \varepsilon_s, \varepsilon_d') - \frac{f(\varepsilon_s) - f(\varepsilon_d')}{\varepsilon_d' - \varepsilon_s}\right] \times \cos[\alpha(\varepsilon_s - \varepsilon_d')] d\varepsilon_d' d\varepsilon_s + \frac{V_{sd} + V_{ds}}{2} N_s N_d \int_{-\infty}^{\infty}\int_{-\infty}^{\infty}\left[L_{ds}(\omega, \varepsilon_d, \varepsilon_s') - \frac{f(\varepsilon_d) - f(\varepsilon_s')}{\varepsilon_s' - \varepsilon_d}\right] \times \cos[\alpha(\varepsilon_d - \varepsilon_s')] d\varepsilon_s' d\varepsilon_d = V_{ss} N_s N_s I_{ss}(\omega, R, T) + V_{dd} N_d N_d I_{dd}(\omega, R, T) + \frac{V_{sd} + V_{ds}}{2} N_s N_d I_{sd}(\omega, R, T) + \frac{V_{sd} + V_{ds}}{2} N_s N_d I_{ds}(\omega, R, T) \quad (9)$$

with $\alpha = \frac{R}{\hbar p_F}$.

The coefficient of $I_{sd}$ takes into account the single particle scattering from $s$ to $d'$, as well as the scattering from $d'$ to $s$. The expression (9) in the limit $s \to 0$ is,

$$I_{ss}(\omega, R, T) = -\pi i \int_{A-\hbar\omega}^{\infty}[1 - 2f(E_{s2})][g(E_s, E_{s2}, A, A)\cos(\alpha\varepsilon_{s2}) - i\sin(\alpha\varepsilon_{s2})]e^{i\alpha\varepsilon_s} dE_s + \pi i \int_{A}^{\infty}[1 - 2f(E_s)][g(E_s, E_{s2}, A, A)\cos(\alpha\varepsilon_s) + i\sin(\alpha\varepsilon_s)]e^{-i\alpha\varepsilon_{s2}} dE_s$$

$$I_{dd}(\omega, R, T) = -\pi i \int_{B-\hbar\omega}^{\infty}[1 - 2f(E_{d2})][g(E_d, E_{d2}, B, B)\cos(\alpha\varepsilon_{d2}) - i\sin(\alpha\varepsilon_{d2})]e^{i\alpha\varepsilon_d} dE_d + \pi i \int_{B}^{\infty}[1 - 2f(E_d)][g(E_d, E_{d2}, B, B)\cos(\alpha\varepsilon_d) + i\sin(\alpha\varepsilon_d)]e^{-i\alpha\varepsilon_{d2}} dE_d$$

$$I_{sd}(\omega, R, T) = -\pi i \int_{A-\hbar\omega}^{\infty}[1 - 2f(E_{s2})][g(E_s, E_{s2}, A, B)\frac{\varepsilon_s}{\varepsilon_s^*}\cos(\alpha\varepsilon_{s2}) - i\sin(\alpha\varepsilon_{s2})]e^{i\alpha\varepsilon_s^*} dE_s + \pi i \int_{A}^{\infty}[1 - 2f(E_s)][g(E_s, E_{s2}, A, B)\frac{\varepsilon_{s2}}{\varepsilon_{s2}^*}\cos(\alpha\varepsilon_s) + i\sin(\alpha\varepsilon_s)]e^{-i\alpha\varepsilon_{s2}^*} dE_s$$

$$I_{ds}(\omega, R, T) = -\pi i \int_{B-\hbar\omega}^{\infty}[1 - 2f(E_{d2})][g(E_d, E_{d2}, A, B)\frac{\varepsilon_d}{\varepsilon_d^*}\cos(\alpha\varepsilon_{d2}) -$$

$i\sin(\alpha\varepsilon_{d2})]\,e^{i\alpha\varepsilon_d^*}dE_d + \pi i\int_B^\infty[1-2f(E_d)][g(E_d,E_{d2},A,B)\frac{\varepsilon_{d2}}{\varepsilon_{d2}^*}\cos(\alpha\varepsilon_d)+i\sin(\alpha\varepsilon_d)]\,e^{-i\alpha\varepsilon_{d2}^*}dE_d$ (10)

with $E_{s2}=E_s+\hbar\omega$, $E_{d2}=E_d+\hbar\omega$

$\varepsilon_s^*=\sqrt{E_s^2-B^2}$, $\varepsilon_{s2}=\sqrt{E_{s2}^2-A^2}$, $\varepsilon_{s2}^*=\sqrt{E_{s2}^2-B^2}$, $\varepsilon_d^*=\sqrt{E_d^2-A^2}$, $\varepsilon_{d2}=\sqrt{E_{d2}^2-B^2}$, $\varepsilon_{d2}^*=\sqrt{E_{d2}^2-A^2}$, and $g(E_1,E_2,\Delta_1,\Delta_2)=\frac{E_1E_2+\Delta_1\Delta_2}{\varepsilon_1\varepsilon_2}$.

When A=B, (10) reduces to equation (3.5) in reference [3].

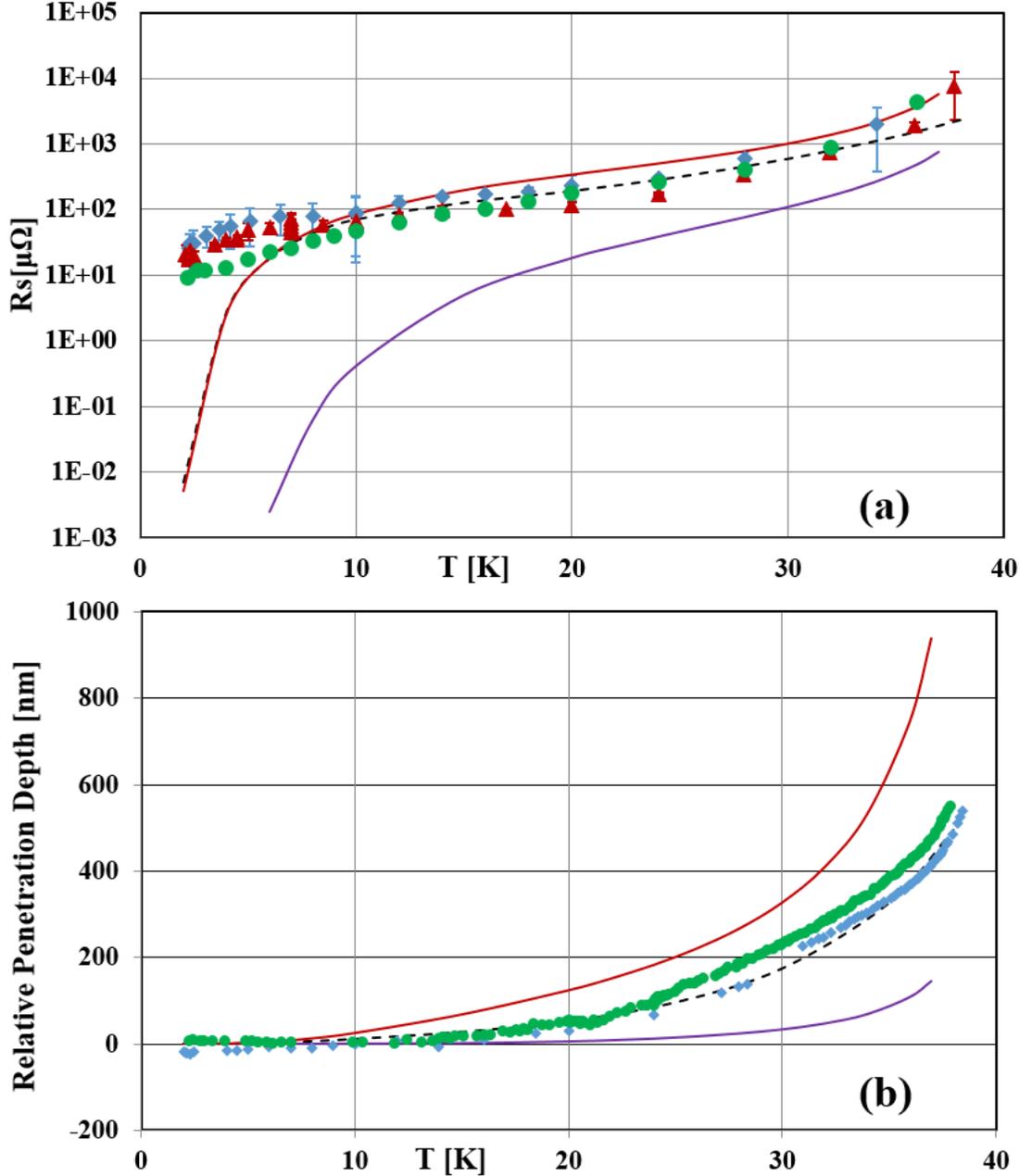

Figure 1. (a) Surface resistance, and (b) relative penetration depth of $MgB_2$. ◆ $MgB_2$-200-I ▲ $MgB_2$-200-II ● $MgB_2$-350, black dotted line the calculation results based on this paper, red line single π-gap results and purple line single σ-gap results.

We use the SRF application of $MgB_2$ at 7.5 GHz range as an example. In this material the large energy gap is much bigger than the small energy gap. In this application $A - \hbar\omega > B$ holds in the temperature range the SRF application is interested in, i.e., from 0 K to 0.1K below the critical temperature.

To obtain the surface impedance, we use the expressions in reference [21],

$$Z = i\pi\omega\mu_0 \{\int_0^\infty ln[1 + K(p)/p^2]dp\}^{-1} \quad (11)$$

with $K(p) = \frac{-3}{4\pi\hbar v_F \lambda_L^2(0)} \int_0^\infty \int_{-1}^1 e^{ipRu} e^{-\frac{R}{l}} (1 - u^2) \times I(\omega, R, T) du dR$ (12)

with $\lambda_L$ London penetration depth, and $K(p)$ the component that connects the Fourier components of current $j(p)$ and vector potential $A(p)$ by:

$$\boldsymbol{J}(\boldsymbol{p}) = -\frac{c}{4\pi} K(p) A(\boldsymbol{p})$$

One now obtains an analytical expression for the surface impedance by incorporating (9) and (10) into (12), then into (11).

A Mathematica™ program has been developed to calculate this integral. In calculations below we use the following parameters as a reference: $A_0$ = 7.1 meV, $B_0$ = 2.3 meV [10], $T_c$ = 39.5 K [9], $l$ = 40 nm, $\lambda_L$ = 100 nm [9], Debye temperature = 884 K [22] and Fermi velocity = 4.7×10$^5$ m/s [23], with outside conditions $T$ = 2 K and frequency at 7.5 GHz. The interaction factors are $V_{sd}$ = 0.119, $V_{ds}$ = 0.09, $V_{ss}$ = 0.81, $V_{dd}$ = 0.285, [24] and $N_s$ = 1, $N_d$ = 1.3 [25, 26].

Using the parameters above, the surface impedance of $MgB_2$ is calculated and compared with the experimental results in [4], with two samples having a 200 nm thick $MgB_2$ layer on sapphire ($MgB_2$-200-I and $MgB_2$-200-II), and one sample with 350 nm thick $MgB_2$ on sapphire ($MgB_2$-350). The surface resistance data and theory are shown in Figure 1(a). Since the SIC system measures the changes of penetration depth (relative penetration depth), the measured and simulated results of relative penetration depth are shown in Figure 1(b). The surface resistance data and theory agree above ~7 K. Below that temperature, the experimental surface resistance levels out and appears dominated by some residual resistance mechanism not included in the theory.

For comparison, the calculation results of the two-band analysis are compared with the results for single band superconductors. With all other parameters the same as the above, surface impedance at different energy gaps, at 7.1 meV (σ-gap), and 2.3 meV (π-gap) were calculated; results are shown in Figure 1. As expected, the surface impedance of a two-band superconductor is greater than a single band superconductor with a σ-gap, and is smaller than that with a π-gap. The surface resistance is clearly dominated by the smaller π-gap [26, 27]. The penetration depth can also be fitted by an effective gap [27], shown in Figure 2 in reference [4] at 3.7 meV.

### III. Conclusion and discussion

With recent activities that are pushing the performance of Nb SRF cavities to their theoretical limits, the necessity of finding alternative materials for SRF application becomes more important. $MgB_2$, as the conventional superconductor with highest critical temperature, is a favorite candidate. By applying the theory of a two-band superconductor to the MB theory, we calculated the surface impedance of $MgB_2$. The results agrees well with previously published experimental results.

Among all type I superconductors, $MgB_2$ has the highest critical temperature and thus is attractive for higher temperature SRF applications with corresponding lower cryogenic power demands, perhaps within range of closed-cycle cryocooling systems. In the sample measurement reported in Figure 1, the residual resistance dominates at temperatures below 7 K. The mechanism(s) for these losses are yet unknown. Perhaps recent activities on Nb cavities [28, 29] could be studied with $MgB_2$ thin films to further elucidate and suppress the residual resistance. Significant R&D is needed to make $MgB_2$ suitable for SRF applications at viable accelerating field gradients. First beneficial applications in accelerator systems could perhaps be in low-field regions that are at intermediate temperature, e.g. RF couplers, beamlines, or beamline bellows, which rely on bulk conduction cooling or use return liquid or gas helium from other SRF or superconducting magnet components.

### ACKNOWLEDGEMENT

This work is supported by Brookhaven Science Associates, LLC under U.S. DOE contract No. DE-AC02-98CH10886, and by Jefferson Science Associates, LLC under U.S. DOE Contract No. DE-AC05-06OR23177. The U.S. Government retains a non-exclusive, paid-up, irrevocable, world-wide